\documentclass[a4paper,11pt]{article}

\usepackage{aas_macros}
\usepackage{physics}
\usepackage[margin=1.95cm]{geometry}
\usepackage{graphicx}
\usepackage{subcaption}
\usepackage{amsmath,amssymb}
\numberwithin{equation}{section}
\usepackage{hyperref}
\usepackage{multirow}

\usepackage{rotating}

\usepackage[dvipsnames]{xcolor}

\hypersetup{
    colorlinks=true,
    urlcolor=ForestGreen,
    linkcolor=blue,
    citecolor=magenta}
    
\newcommand{\emsp}{\mspace{12mu}}
\newcommand{\redshift}{{z}}

\title{Constraints on Bianchi-I type universe with SH0ES anchored Pantheon+ SNIa data}
\author{Anshul Verma$^1$, Sanjeet K. Patel$^1$, Pavan K. Aluri$^1$, Sukanta Panda$^2$, David F. Mota$^3$}
\date{\today}

\begin{document}

\maketitle

\centerline{$^1$Department of Physics, Indian Institute of Technology (BHU), Varanasi - 221005, India}
\centerline{$^2$Department of Physics, Indian Institute of Science Education and Research, Bhopal - 462066, India}
\centerline{$^3$Institute of Theoretical Astrophysics, University of Oslo, P.O. Box 1029 Blindern, N-0315 Oslo, Norway}

\begin{abstract}
We study the Bianchi-I cosmological model motivated by signals of statistical isotropy violation seen in cosmic microwave background (CMB) observations and others. To that end, we consider various kinds of anisotropic matter that source anisotropy in our model, specifically Cosmic strings, Magnetic fields, Domain walls and Lorentz violation generated magnetic fields. These anisotropic matter sources, taking one at a time, are studied for their co-evolution with standard model (isotropic) sources viz., dust-like (dark/normal) matter, and dark energy modelled as cosmological constant. We constrain the Hubble parameter, density fractions of anisotropic matter, cold dark matter (CDM), and dark energy ($\Lambda$) in a Bianchi-I universe with planar symmetry i.e., which has a global ellipsoidal geometry, and try to find signatures of a cosmic preferred axis if any.
The latest compilation of Type Ia Supernova (SNIa) data from Pantheon+SH0ES collaboration is used in our analysis to obtain constraints on cosmological parameters and any preferred axis for our universe.
In our analysis, we found mild evidence for a cosmic preferred axis. It is interesting to note that this preferred axis lies broadly in the vicinity of other prominent cosmic anisotropy axes reported in the literature from diverse data sets.
Also we find some evidence for non-zero (negative) cosmic shear and eccentricity that characterize different expansion rates in different directions and deviation from an isotropic scale factor respectively.
The energy density fractions of two of the sources considered are found to be non-zero at a
$2\sigma$ confidence level.
To be more conclusive, we require more SNIa host galaxy data for tighter constraints on distance and absolute magnitude calibration which are expected to be available from the future JWST observations and others.

\end{abstract}

\section{Introduction}
Observations of Cosmic Microwave Background (CMB) from COBE, WMAP, and Planck missions~\cite{cobe1996maps,wmap9yrmaps,plk2018maps} led to the current era of `Precision Cosmology'. This naturally paved the way in-turn for testing the fundamental hypothesis of current standard model of cosmology itself, specifically homogeneity and isotropy of our universe. It resulted in finding many instances of isotropy violation in CMB large angular scales indicating a preferred direction for our universe~\cite{wmap7yranom,plk2018isostat,schwarz2016}.
Such violations of statistical isotropy were also seen in other data sets such as Type Ia Supernovae~\cite{schwarz2007sn1a,perivolaropoulos2010sn1a,subir2011sn1a,wiltshire2013sn1a,
appleby2015sn1a,zhao2019sn1a}, alignment of optical polarization vectors of distant Quasars~\cite{hutsemekers1998,hutsemekers2001,jain2004,hutsemekers2014}, radio polarization vector alignments~\cite{birch1982,jain1999,tiwari2013radioalgn}, large scale velocity flows being larger than those predicted in $\Lambda$CDM cosmology~\cite{watkins2009}, etc. A host of other observational challenges to the current standard model of cosmology are also present. See, for example, 
Ref.~\cite{bull2016,abdalla2022,perivolaropoulos2022,aluri2023cp} for a review. 
Unexpected alignments of anisotropy axes seen in such diverse astronomical and cosmological data sets were initially discussed in Ref.~\cite{johnjain2004}. Although, we are in an era of unprecedented precision in terms of cosmological data that is available, and the standard cosmological model based on `Cosmological principle' explaining a variety of observational data, we still have these challenges or anomalies mentioned above that needs to be resolved for a better understanding of our universe.

Many of these anomalies suggest re-examing the notion of `isotropy' via the Cosmological principle. Such anisotropy axes may be indicative of different expansion rates in different directions of our universe. In order to explain such anisotropy, we consider a global geometry using some metric that is non-trivial and yet has minimal modifications to the Friedmann-Lema\^itre-Robertson-Walker (FLRW) background. There exists such a class of anisotropic, but homogeneous models called Bianchi-type cosmological models~\cite{maccallumellis1970}. The simplest among them is the Bianchi-I model whose invariant space-time line element is given by
\begin{equation}
ds^2 = c^2 dt^2 - a(t)^2 dx^2  - b(t)^2 dy^2 - c(t)^2 dz^2\,,
\label{eq:bianchi1}
\end{equation}
where `$a(t)$', `$b(t)$' and `$c(t)$' are the scale factors characterizing expansion of the universe along the three comoving coordinates $x$, $y$ and $z$ respectively. Further, the constant factor `$c$' is the speed of light in vacuum and `$t$' is the cosmic time coordinate. Hereafter, we work in units of $c=1$. This difference in cosmic expansion in different directions leads to an axis of anisotropy for the universe. Given the nature of physical sources which will be studied in the present work that can induce anisotropy, we consider Eq.~(\ref{eq:bianchi1}) with a residual planar symmetry taking $a(t)=b(t)$ and the $z$-axis normal to the plane denoting universe's preferred direction, which will be constrained using observational data later in this paper.

While we change the left hand side of Einstein's field equations by assuming a different metric for our background space-time, we also need to take care of the right hand side by choosing appropriate source terms via the energy-momentum tensor. The question is what kind of matter can be added as a source of anisotropy (if there is any) in the universe? During inflation, the anisotropic shear in the universe decreases and becomes isotropically negligible at last scattering surface~\cite{emir2007}. To address any non-trivial shear at present times, there has to be some source of anisotropy after that period. This may happen if we have an anisotropic dark energy or any other form of anisotropic matter source. In this paper, we investigate the latter case with four types of anisotropic sources : Cosmic Strings (CS), Domain Walls (DW), Lorentz Violation generated Magnetic Fields (LVMF), and Magnetic Fields (MF)~\cite{barrow1997,barera2004}. It was shown in Ref.~\cite{campanelli2006} that power suppression at low multipoles can be explained by assuming magnetic fields. Large scale magnetic fields could have been produced during cosmic inflation due to a Lorentz-violating term in the  photon sector~\cite{campanelli2009}. Anisotropic dark energy as a possible source was discussed, for example, in Ref.~\cite{koivisto2008}.

We note that this broad class of homogeneous but anisotropic Bianchi models, which have an FLRW space-time as a limiting case, were studied in the context of CMB for the kind of patterns they can induce in CMB temperature and polarization signals (see for example Ref.~\cite{coles2011}). Though attempts to address CMB `anomalies' using various Bianchi model templates were partly successful, it also lead to cosmological parameter inconsistencies with the standard model (see for example Refs.~\cite{jaffe2006,saadeh2016}).
Similar attempts were made to study any signatures of a preferred direction for our universe using SNIa data, based on an anisotropic cosmological model or empirically probing the data for such violations, with differing results (see for example Refs.~\cite{schwarz2007sn1a,subir2011sn1a,zhao2019sn1a,koivisto2008,campanelli2011,rahman2022}).

Rest of the paper is organized as follows. In the next section we briefly discuss Einstein's field equations in cosmic time. Also presented is the distance modulus - redshift relation for a Bianchi-I space-time along with the fractional density and Hubble parameter evolution equations to be solved for obtaining constraints later on.
In section~\ref{sec:SNIa-data-method}, the observational data used and the likelihood analysis to constrain the model parameters are explained in brief. Then, in section~\ref{sec:results}, we present our results and analysis i.e., constraints obtained on various fractional energy densities of (an)isotropic sources considered for this work and the presence of any cosmic preferred axis. These are followed by section~\ref{sec:concl} where we discuss and summarize our findings.

\section{Bianchi-I Universe}
\label{sec:bianchi1}

\subsection{Planar Bianchi-I space-time and Sources of anisotropy}
\label{sec:bianchi1planar}
Given the hints for \emph{anisotropy} from diverse astronomical data, we should consider a metric that generalizes flat FLRW metric and may explain the cosmological skewness~\cite{barrow1997,barera2004,campanelli2006}.
To that end, we study the Bianchi-I space-time given by Eq.~(\ref{eq:bianchi1}) but with
residual planar symmetry described by the line element,
\begin{equation}
ds^2 = dt^2 - a(t)^2 (dx^2  + dy^2) - b(t)^2 dz^2\,,
\label{eq:bianchi1plnr}
\end{equation}
where `$a(t)$' and `$b(t)$' are now the scale factors perpendicular to and along the axis of anisotropy. By virtue of its geometry, it is also called Eccentric or Ellipsoidal Universe~\cite{barera2004,campanelli2006}.

The diagonal anisotropic energy-momentum tensor for this metric is then taken to be,
\begin{equation}
    T^{\mu}_{\emsp \nu} = {\rm diag}(\rho,-p_a,-p_a,-p_b)\,,
\label{eq:b1emtensor}
\end{equation}
where, $p_a$ = $w_a\rho$ and $p_b$ = $w_b\rho$ are the anisotropic pressure terms in the plane normal to and along the aniostropy axis. As mentioned earlier, we consider four types of anisotropic matter sources in this paper. First we consider two types of topological defects viz., cosmic strings (CS) and domain walls (DW) as a form of anisotropic matter. 
Then we also consider a magnetic field generated due to a Lorentz-violating term in the photon sector, abbreviated as LVMF, as another anisotropic source. Finally the well known configuration of a uniform magnetic field is also considered.
Their anisotropic equations of state are summarized in Table~\ref{tab:am-eos}~\cite{barrow1997,barera2004,campanelli2009,aluri2013sn1a}.
Note that because of the symmetry one assumes in Eq.~(\ref{eq:bianchi1})
i.e., depending on our choice of geometry being prolate or oblate
($a(t)=b(t)$ or $b(t)=c(t)$), the equation of state parameters, $w_a$ and $w_b$, may get
interchanged accordingly for all anisotropic sources. Also, the convention used to
define the energy-momentum tensor, Eq.~(\ref{eq:b1emtensor}),
or the signature of the metric in Eq.~(\ref{eq:bianchi1plnr}),
will further result in a change of sign for $w_a$ and $w_b$ listed in Table~\ref{tab:am-eos}.
		 
Among these, the cosmic string and domain wall distributions can grow gradually anisotropic with time~\cite{vilenkin1988}. The origin of all these sources is primordial in the sense that all of these were produced through broken discrete symmetries during phase transitions in the early universe~\cite{kolbturner1990}. We study the anisotropic co-evolution of these sources with dust-like (normal+dark) matter and dark energy represented by a cosmological constant, $\Lambda$. Corresponding to our choice of energy-momentum tensor, we have two different equations of state $w_a$ along any direction within the plane of symmetry and $w_b$ along the normal direction to that plane.

The necessary mathematical relations following Einstein's field equations, that are required to constrain the Ellipsoidal model being studied were earlier presented in Refs.~\cite{koivisto2008,aluri2013sn1a}. However, we breifly summarize them consistently for the form of planar Bianchi-I metric given in Eq.~(\ref{eq:bianchi1plnr}).

\begin{table}
\begin{center}
    \begin{tabular}{ | l || c | c | }
    \hline
    Anisotropic Source& $w_a$ & $w_b$\\ 
    \hline
    Cosmic Strings& 0 & -1\\
    Domain Walls& -1 & 0\\
    LVMF & 0 & 1\\
    MF & 1 & -1\\
    \hline
    \end{tabular}
\end{center}
\caption{Anisotropic sources used in this paper and their anisotropic equation of
         state parameter values are given here.}
\label{tab:am-eos}
\end{table}

\subsection{Einstein's field equations}
The Einstein's field equations following our choice of metric given by Eq.~(\ref{eq:bianchi1plnr}) and anisotropic stress-energy tensor given by Eq.~(\ref{eq:b1emtensor}) can be found to be,
\begin{eqnarray}
\label{eq:b1efe}
    H_a^2  + 2H_a H_b &=& 8\pi G\rho\,, \nonumber \\
    \dot{H_a} + \dot{H_b} + H_a^2 + H_b^2 + H_aH_b &=& -8\pi Gp_a\,, \\
    2\dot{H_a} + 3H_a^2 &=& -8\pi Gp_b\,, \nonumber 
\end{eqnarray}
where $H_a = \dot{a}/a$ and $H_b=\dot{b}/b$ are the rates of expansion or Hubble parameters
in and normal to the plane of residual symmetry. Here, $\dot{a} = da/dt$ and $\dot{b}=db/dt$.
The continuity equation, $\nabla_\mu T^\mu_{\emsp \nu}=0$, leads to finding,
\begin{equation}
    \dot{\rho} + (2H_a + H_b)\rho + 2H_a p_a +H_bp_b = 0\,.
\label{eq:b1conteq}
\end{equation}
We set $8\pi G=1$ for the rest of this paper. The expansion rates $H_a$ and $H_b$ can
be more conveniently expressed in terms of average Hubble parameter and
differential expansion rate - also called the shear parameter - as,
\begin{equation}
H= (2H_a + H_b)/3 \quad {\rm and} \quad \Sigma = (H_a - H_b)/\sqrt{3}\,,
\end{equation}
respectively.
In terms of these new variables, Eq.~(\ref{eq:b1efe}) and (\ref{eq:b1conteq}) can be
rewritten with respect to cosmic time `$t$' as:
\begin{eqnarray}
\label{eq:b1efe-new}
    \frac{dH}{dt} &=& -H^2 - {\frac{2}{3}}\Sigma^2 - {\frac{1}{6}}(\rho + 2p_a + p_b)\,, \nonumber \\
    \frac{d\Sigma}{dt} &=& -3H\Sigma + \frac{1}{\sqrt{3}}(p_a - p_b)\,, \\
    \frac{d\rho}{dt} &=& -3H(\rho + \frac{2p_a + p_b}{3}) - \frac{2\Sigma}{\sqrt{3}}(p_a - p_b)\,, \nonumber
\end{eqnarray}
along with a constraint equation given by,
\begin{equation}
    H^2 = \frac{\rho}{3} + \frac{\Sigma^2}{3}\,.
\label{eq:b1consteqn}
\end{equation}
Note that the isotropic non-interacting sources viz., dust-like matter and a cosmological
constant are automatically described by the anisotropic energy-momentum tensor given in Eq.~(\ref{eq:b1emtensor}), where $p_a=p_b$ or equivalently $w_a=w_b$.

\subsection{Evolution equations and Luminosity distance}
\label{sec:efe-Dl}

In order to constraint various parameters characterizing our Bianchi-I model, we employ the latest Type Ia Supernova (SNIa) data available that will be described in the next section.
Here we present the theoretical luminosoty distance ($d_L$) versus redshift ($\redshift$)
relation that is dependent on various cosmological parameters of our model
and the evolution equations of these parameters to be solved in conjugation.
First we define the mean scale factor and the eccentricity parameter (characterising
deviation from isotropic expansion) in a Bianchi-I model as,
\begin{equation}
A = (a^2 b)^{1/3} \quad {\rm and } \quad e^2 = 1-\frac{b^2}{a^2}\,.
\label{eq:mean-scalefac-ecc}
\end{equation}
This definition of mean scale factor can be seen to be easily reconciled with
our definition of mean Hubble paramter, $H$, introduced earlier as,
\begin{equation}
H = \frac{\dot{A}}{A} = \frac{2H_a+H_b}{3}\,.
\end{equation}
Further, it is convenient to define some dimensionless variables as follows.
First, using the constraint equation, Eq.~(\ref{eq:b1consteqn}), we define dimensionless
energy density paramter $\Omega=\rho/(3H^2)$ (in units of $c=8\pi G = 1$) and
expansion normalized dimensionless shear parameter $\sigma=\Sigma/(\sqrt{3}H)$.

Here we note that the energy density $\rho$ in Eq.~(\ref{eq:b1consteqn}), and consequently
$\Omega$, is comprised of the two isotropic sources viz., (ordinary+dark) matter
denoted by `$\Omega_{IM}$' and the dark energy `$\Omega_\Lambda$' due to cosmological constant term, and the anisotropic sources denoted by $\Omega_{AM}$. As listed in Table~{\ref{tab:am-eos}}, we consider four types of anisotropic matter i.e., cosmic strings, domain walls, magnetic fields and Lorentz violation generated magnetic fields taking one at a time, along with isotropic standard model components. By defining critical energy denisty as
$\rho_{c,0}$ or simply $\rho_0=3H^2_0$, where $H_0$ is the value of mean Hubble parameter at current time, `$t_0$', we can write the constraint equation at present as,
\begin{equation}
\Omega_{AM,0}+\Omega_{IM,0}+\Omega_{\Lambda,0} + \sigma^2_0 = 1\,,
\label{eq:b1-constraint-eqn}
\end{equation}
where
\begin{equation}
\Omega_{AM,0} = \rho_{AM,0}/\rho_0\,, \quad \Omega_{IM,0} = \rho_{IM,0}/\rho_0\,, \quad {\rm and}
\quad \Omega_{\Lambda,0} = \rho_\Lambda/\rho_0\,,
\end{equation}
are the fractional energy densities of various sources at current time, $t_0$.
Hereafter we drop the subscript `$0$' in the fractional energy densities for brevity, and simply denote them as $\Omega_{AM}$, $\Omega_{IM}$, $\Omega_\Lambda$. Nevertheless their usage should be clear from the context, whether they denote energy densities at current time having fixed values or time dependent variables. Finally, a dimensionless time variable is defined as $\tau=\ln(A)$. Thus we have $H=d\tau/dt$.

The direction dependent luminosity distance ($d_L$) of an SNIa object, observed in the direction $\hat{n}=(\theta$, $\phi$) in a Bianchi-I universe is given by~\cite{koivisto2008},
\begin{equation}
 d_L(\hat{n}) = c(1 + \redshift_{\rm hel})\int_{A(\redshift^*)}^1{\frac{dA}{{A^2}H}}{\frac{(1 - e^2)^\frac{1}{6}}{(1 - e^2\cos^2\alpha)^\frac{1}{2}}}\,,
\label{eq:b1lumdist}
\end{equation}
where $c$ is the speed of light,  $\alpha = \cos^{-1}(\hat{\lambda}\cdot\hat{n})$ is the angle
between cosmic preferred axis `$\hat{\lambda}$' (i.e., the $z$-axis of Eq.~{\ref{eq:bianchi1plnr}}) and the location of an SNIa at `$\hat{n}$' in the sky.
Further, here, `$z_{\rm hel}$' and `$z^*$' are the heliocentric and cosmological redshifts of an SNIa, respectively~\cite{davis2011}.

To solve for the intergral in the above expression for `$d_L$', the complete set of evolution
equations (following Eq.~({\ref{eq:b1efe-new}}) and Eq~(\ref{eq:mean-scalefac-ecc})) in terms of dimensionless variables, just defined, can be found to be~\cite{aluri2013sn1a},
\begin{eqnarray}
\label{eq:b1evoleq}
    \frac{h'}{h} &=& -\frac{3}{2}(1+\sigma^2 +\bar{w}\Omega_{AM} - \Omega_\Lambda)\,, \nonumber\\
    (e^2)' &=& 6\sigma(1-e^2)\,, \nonumber \\
    \sigma' &=& -\frac{3}{2}[\sigma(1-\sigma^2) - (\frac{2}{3}\delta_w + \bar{w}\sigma)\Omega_{AM} + \sigma\Omega_\Lambda]\,, \\
    \Omega_{AM}' &=& -3\Omega_{AM}(\bar{w} + \frac{2}{3}{\delta_w}\sigma - \bar{w}\Omega_{AM} + \Omega_\Lambda - \sigma^2)\,, \nonumber \\
    \Omega_{IM}' &=& 3\Omega_{IM}(\bar{w}\Omega_{AM} - \Omega_\Lambda + \sigma^2)\,, \nonumber\\
    \Omega_\Lambda' &=& 3\Omega_\Lambda(1+ \bar{w}\Omega_{AM} - \Omega_\Lambda + \sigma^2)\,, \nonumber
\end{eqnarray}
where, for some parameter `$X$', its prime denotes a derivative with respect to the new time variable, $\tau$, i.e., $X'=dX/d\tau$\footnote{We note that there was a typographical error in Eq.~7.9 of Ref.~\cite{aluri2013sn1a} for $e^2$ - an extra factor of $1/A$ - that is correctly presented here.}, $\bar{w}=(2w_a+w_b)/3$ and $\delta_w = w_a-w_b$.
Here we dont need to solve for, say, $\Omega_\Lambda$ which is automatically fixed by the constraint equation $\Omega_\Lambda = 1 - \Omega_{IM}-\Omega_{AM}-\sigma^2$ at any time. Further, $h=H/(100~km/s/Mpc)$ in the preceeding set of equations is the dimensionless Hubble parameter.

Since we are provided with redshift ($\redshift^*$) of an object, these evolution equations can be solved for any of the cosmological parameters `$X$' with respect to redshift
`$\redshift$' as,
\begin{equation}
\frac{dX}{d\redshift} = \frac{dA}{d\redshift}\frac{d\tau}{dA}\frac{dX}{d\tau} = \frac{1}{A} \frac{dA}{d\redshift} X'\,,
\end{equation}
where we used the relation $d\tau/dA=1/A$. Also, $dA/d\redshift$ can be found to be given by,
\begin{eqnarray}
\label{eq:dAdz}
   \frac{dA}{d\redshift} &=& \frac{-f_1}{(1+\redshift)(1+\redshift-{f_1}{f_2}\frac{\sigma}{A})}\,, \nonumber \\
    f_1 &=& \frac{(1 - e^2\sin^2\alpha)^\frac{1}{2}}{(1 - e^2)^\frac{1}{3}}\,,\\
    f_2 &=& \frac{2+(e^2 - 3)\sin^2{\alpha}}{1 - e^2\sin^2{\alpha}}\,. \nonumber
\end{eqnarray}
Therefore, the evolution equations are solved from $\redshift=0$ (corresponding to the initial condition $A_0=1$ at current time $t=t_0$) up to $\redshift=\redshift^*$ in order to obtain constraints on our model parameters.
The above equation for $dA/dz$ is used to solve the integral in Eq.~(\ref{eq:b1lumdist})
in terms of redshift `$z$' rather than the mean scale factor `$A$'.
The set of cosmological parameters to be constrained in the anisotropic Bianchi-I model are
$
\{h_0, e^2_0, \sigma_0, \Omega_{AM}, \Omega_{IM}, \Omega_\Lambda, \hat{\lambda} = (l_a,b_a)\}
$,
where we consider each of the anisotropic matter sources (given in Table~\ref{tab:am-eos}) one at a time, along with the isotropic sources. As noted earlier, $\Omega_\Lambda$ is not constrained independently but, by default, fixed via the constraint equation given by Eq.~(\ref{eq:b1-constraint-eqn}). The cosmic anisotropy axis in our model denoted by
$\hat{\lambda} = (l_a,b_a)$ will be estimated in galactic coordinates.

Constraints thus obtained on our anisotropic model will be compared with the standard flat
$\Lambda$CDM model, for which the luminosity distance in given by,
\begin{equation}
d_L = \frac{c(1 + \redshift_{\rm hel})}{H_0}\int_0^{\redshift^*}{\frac{d\redshift}{\sqrt{\Omega_{IM}(1 + \redshift)^3 + \Omega_\Lambda}}},
\label{eq:flrw-lumdist}
\end{equation}
where $\Omega_\Lambda = 1-\Omega_{IM}=1-\Omega_{M}$, and $\Omega_{M}$ is the current (dust-like dark/visible) matter density fraction in the universe in the standard model. Thus we have two cosmological parameters viz., $H_0$ (or equivalently $h_0$) and
$\Omega_{M}$ (or $\Omega_\Lambda$), to be constrained in the standard concordance model,
which will be our reference model to compare with.

\section{Data sets used and Methodology}
\label{sec:SNIa-data-method}

\subsection{The Pantheon+ SNIa data}
\label{sec:SNIa-data}
The Pantheon+ SNIa data set\footnote{\url{https://github.com/PantheonPlusSH0ES/DataRelease}} is a compilation of 1701 light curves of about 1550 unique spectroscopically confirmed SNIa\footnote{To be specific we found only 1543 distinct SNIa in the data made publicly available.}~\cite{scolnic2022}, superseding the previous Pantheon dataset that contained 1048 spectroscopically confirmed SNIa~\cite{scolnic2018}. The light curves of rest of the SNIa i.e., other than about 1550 out of 1701 are either the same SNIa measured by different surveys \emph{or} are ``SN siblings'' meaning Supernovae (SNe) found in the same host galaxy. Major improvements in the latest dataset are the addition of new SNe, and peculiar velocity and host galaxy bias corrections in the SNe covariance matrix.

Only the best fitting Supernovae to \emph{Type Ia} which includes class identification,
measurement time, convergence of the fit and other sample cuts, were included in Pantheon+
data set. For example, the PS1 survey includes three level quality cuts namely Initial criteria, First-pass criteria and the Final criteria~\cite{ps12014}. Similarly, SNe from other surveys were also subjected to several quality checks to confirm that they unambiguously fall under `Type Ia' category as physics of them is presumably well understood, although there exists debate in literature on their progenitors~\cite{progenitorsMaeda2018,progenitorsShen2019}.

\begin{figure}[t]
    \centering
    \includegraphics[height=4.4cm]{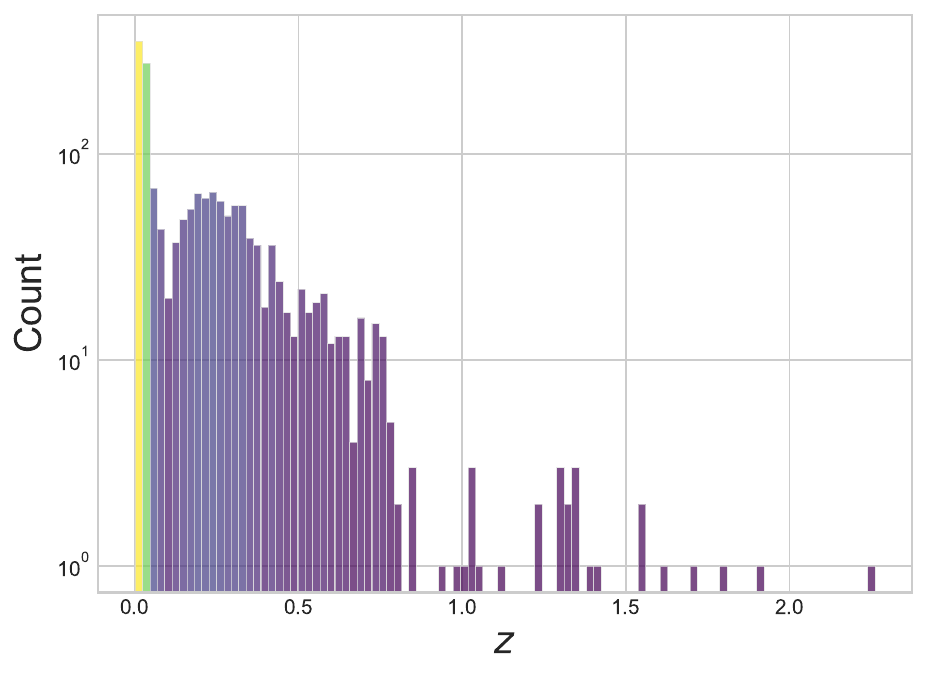}
    ~
    \includegraphics[height=4.6cm]{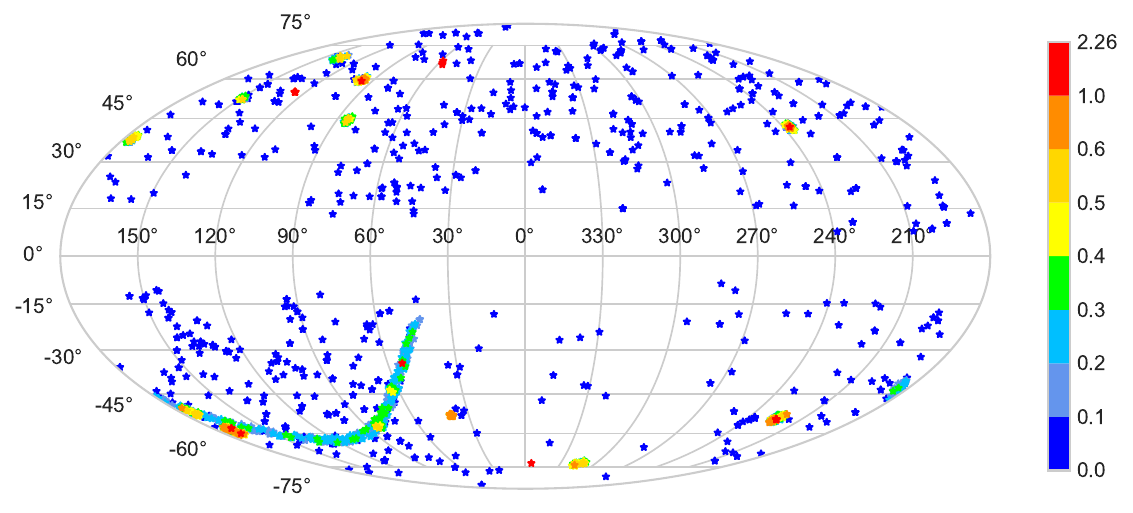}
    \caption{\emph{Left :} Number count histogram of SNIa in Pantheon+ data with
    		 respect to redshift in 100 bins.
    		 \emph{Right :} Sky distribution of Pantheon+ SNIa.}
    \label{fig:pantheon+_dist}
\end{figure}

In the \emph{left} panel of Fig.~[\ref{fig:pantheon+_dist}], we exhibit the redshift distribution of the Pantheon+ compilation of Type Ia Supernaovae that we used in our present study. In the \emph{right} panel of the same figure, we have shown the sky distribution of SNIa that make up the Pantheon+ sample, and the accompanying colour bar denotes the redshift of each SNIa in that sample. As is obvious from the figure, the distribution of SNIa
in Pantheon+ sample is quite inhomogeneous with respect to redshift ($\redshift$) as well as spatial locations ($\hat{n}$).

To know the distance of a high-$z$ SNIa, one requires complete information of the `distance ladder'~\cite{rowanrobinson1985,webb1999,degrijs2019} from three rungs : Cepheid time period and parallax measurements from the `first-rung', independent distance calibration of SNIa-Cephied host galaxies of the `second-rung' (galaxies with both SNIa and Cepheid i.e., both objects having a common host galaxy also called `anchors'), and finally high-$z$ (Hubble flow) SNIa observations of the `third-rung'. Note that the independent host galaxy distances from the second rung, which are calibrated utilizing Cepheid
period-luminosity information and parallaxes from the first rung, are of utmost importance in terms of removing the degeneracy between the Hubble parameter `$H_0$' and the SNIa absolute magnitude `$M_0$' when performing the cosmological parameter estimation. `Pantheon+SH0ES'
collaboration used 77 light curves of 42 independent SNIa in 37 hosts and 277 high quality Hubble flow SNIa to constrain $H_0$~\cite{riess2022}. In this paper, we follow the same approach as taken by `Pantheon+'~\cite{brout2022} to constrain our Bianchi-I model.

\subsection{Likelihood analysis}
\label{sec:SNIa-method}
Here, we describe the likelihood function ($\mathcal{L}$) that is dependent on various SNIa fit parameters including cosmological parameters of our interest. The likelihood function is minimized by sampling various parameters that it depends on using an MCMC method.

From the data, one is provided with, among others, the redshift `$\redshift^*$' and distance modulus `$\mu$' for all SNIa including the anchors. Supernova cosmology relies on relating the luminosity distance, $d_L$, which is dependent on cosmological parameters, ${\boldsymbol \theta}=\{h_0, \Omega_{IM}, e^2_0, \sigma_0, \Omega_{AM},(l_a,b_a)\}$ in our case, with the \emph{distance modulus} as,
\begin{equation}
\mu^{\rm th} = 5\log_{10}\left( \frac{d_L({\boldsymbol\theta},\redshift)}{10~pc} \right) = m_{\rm B} - M_0\,. 
\label{eq:distmod}
\end{equation}
where `$m_{\rm B}$' and `$M_0$' are the apparent and absolute B-band magnitudes of a Supernova respectively. It is assumed that `$M_0$' is same for all SNIa.

In order to obtain constrains, we minimize the $\chi^2({\boldsymbol \Theta})$
($=-\ln(\mathcal{L({\boldsymbol \Theta})})$) function that is dependent on the fitting parameters collectively denote by `${\boldsymbol \Theta}$' that includes the cosmological
parameters `${\boldsymbol \theta}$' also mentioned above.
One can form the $\chi^2$ function following the quadrature approach by defining it as,
\begin{equation}
   \chi^2({\boldsymbol \Theta}) = \sum_{i={\rm SNIa}}\frac{(\mu_i^{\rm obs} -
   								\mu_i^{\rm th})^2}{\delta{\mu_i}^2},
\label{eq:chi2-diag}
\end{equation}
when there are no correlations between Supernovae.
But, there are correlations between these SNIa that are calibrated simultaneously due to both
statistical and systematic uncertainties in SNIa light curve fitting - for a given Supernova
as well as among various SNIa. Therefore, we use the full covariance matrix provided as part of Pantheon+ data release~\cite{scolnic2022}.

Thus to find constraints on our model, we define the $\chi^2$ function as,
\begin{equation}
\chi^2({\boldsymbol \Theta}) = \Delta{\boldsymbol \mu}^T ({{\bf C}_{\rm stat+sys}^{\rm SN} + {\bf C}_{\rm stat+sys}^{\rm Cepheid}})^{-1} \Delta{\boldsymbol \mu}\,,
\label{eq:chi2-cov}
\end{equation}
which is minimized with respect to the parameters `${\boldsymbol \Theta}$'.
Further, $\Delta{\boldsymbol \mu}=(\Delta \mu_1 \,\, \Delta \mu_2 \,\, \hdots \,\, \Delta \mu_n)^T$ is a column vector where `$n$' is the number of SNIa in the sample, and $\Delta \mu_i$ is defined as,
\begin{equation}
    \Delta \mu_i =
    \left\{
    \begin{array}{l l l}
    \mu_i^{\rm SNIa} - \mu_i^{\rm Cepheid} & = m^i_{\rm B, corr} - M_0 - \mu_i^{\rm Cepheid}, & \text{if } i \in \text{Cepheid hosts}\,, \\
    \mu_i^{\rm SNIa} - \mu_i^{\rm th} & = m^i_{\rm B, corr} - M_0 - \mu_i^{\rm th}({\boldsymbol \theta},\redshift_i^*), & \text{otherwise}\,.
    \end{array}
	\right.
\label{eq:mu-sn-ceph-host}
\end{equation}
Here $m^i_{\rm B, corr} = m^i_{\rm B} +\alpha x_1 - \beta {\sf c}$ is the \emph{corrected}
apparent B-band magnitude of the $i^{th}$ SNIa, corrected for colour ({\sf c}) and stretch factors ($x_1$) of an SNIa observed light curve adjusted with the corresponding (global) parameters `$\alpha$' and `$\beta$'.
So we will be using the `\verb+m_b_corr+' column for $m^i_{\rm B, corr}$ for SNIa and
`\verb+CEPH_DIST+' for $\mu_i^{\rm Cepheid}$ from the data provided by Pantheon+
collaboration. This `either-or' condition (that is sort of an {\tt if} condition) where Cepheid data is available in an SNIa host galaxy is invoked using the column `\verb+IS_CALIBRATOR+' that contains a `1' or `0' value that is also provided in the Pantheon+ data table. We can also use the uncorrected apparent magnitude $m_{\rm B}^i$ by using the column `\verb+mB+' along with `\verb+x1+' and `\verb+c+' columns for stretch ($x_1$) and colour (${\sf c}$) factors, etc., from the data and also obtain constrains on $\alpha$ and $\beta$ along with the cosmological parameters in this model. However, we proceed with the corrected B-band apparent magnitudes as provided in the data.

Then, we will be using `\verb+zHD+' column from the data table for the redshift of an SNIa, $z_i^*$, that are corrected for peculiar velocity effects including the CMB kinematic dipole.
Further, ${\bf C}_{\rm stat+sys}^{\rm SN}$ denotes the SNe covariance matrix and
${\bf C}_{\rm stat+sys}^{\rm Cepheid}$ corresponds to the SH0ES Cepheid host-distance covariance matrix, that are also provided as part of Pantheon+ data release. The distance
modulus formula can be further simplified to be,
\begin{equation}
\mu^{\rm th}_i = 5\log_{10}\left({\tilde{d}_L({\boldsymbol\theta},\redshift^*_i)}\right) + \mu_0, 
\label{eq:distmod-dimless}
\end{equation}
where $\tilde{d}_L$ is now dimensionless and the constant offset term,
\begin{equation}
\mu_0= 5\log_{10}\left( \frac{c}{100~km/s/Mpc}~ \frac{1}{10~pc} \right) \approx 42.384\,,
\end{equation}
arises as a result of dimensionful parameters in Eq.~(\ref{eq:b1lumdist}) or
Eq.~(\ref{eq:flrw-lumdist}) viz., the speed of light `$c$' (in $km/s$) and the dimensionful
part of Hubble parameter $H=h\times (100~km/s/Mpc)$, and the $10~pc$ term in the denominator
of Eq.~(\ref{eq:distmod}).

Therefore, in a Bianchi-I universe, the dimensionless luminosity distance is given by ,
\begin{equation}
 \tilde{d}_L(\hat{n}) = (1 + \redshift_{\rm hel})\int_{A(\redshift^*)}^1{\frac{dA}{{A^2}h}}{\frac{(1 - e^2)^\frac{1}{6}}{(1 - e^2\cos^2\alpha)^\frac{1}{2}}}\,,
 \label{eq:b1lumdist-dimless}
\end{equation}
following Eq.~(\ref{eq:b1lumdist}). For the standard flat $\Lambda$CDM cosmological model, it is given by,
\begin{equation}
\tilde{d}_L = \frac{(1 + \redshift_{\rm hel})}{h_0}\int_0^{\redshift^*}{\frac{d\redshift}{\sqrt{\Omega_{IM}(1 + \redshift)^3 + \Omega_\Lambda}}}\,,
\label{eq:flrw-lumdist-dimless}
\end{equation}
following Eq.~(\ref{eq:flrw-lumdist}).

An interesting point to note here is that, in the standard $\Lambda$CDM model, the Hubble parameter ($H_0$ or $h_0$) and the absolute SNIa magnitude ($M_0$) are degenerate with each other. One can see from Eq.~(\ref{eq:distmod}) and (\ref{eq:flrw-lumdist-dimless}) that both form an effective one parameter term given by ``$- M_0 + 5\log_{10}(h_0)$''. Hence they cannot be constrained simultaneously using Supernovae data alone.
But, from Eq.~(\ref{eq:b1lumdist-dimless}), we see that there is no such degeneracy between (mean) Hubble parameter and the absolute magnitude of Supernovae in a Bianchi-I universe, at least in principle depending on the strength of anisotropy. So it may be possible to constrain them independently. However, following Eq.~(\ref{eq:mu-sn-ceph-host}), since we will be using Cepheid data where available to complement the SNIa data, we will be able to break that parameter degeneracy between `$M_0$' and `$h_0$'.
The first part of Eq.~(\ref{eq:mu-sn-ceph-host}) depends on `$M_0$' alone, and the second part depends on the combination `$- M_0 + 5\log_{10}(h_0)$' where the `$h_0$' term appears in via the theoretical distance modulus $\mu_i^{\rm th}({\boldsymbol \theta},\redshift_i^*)$. Thus using the SH0ES anchor (Cepheid) data, we will be able to break that degeneracy between `$M_0$' and `$h_0$' So this degeneracy will no longer be a concern in constraining them simultaneously.

\section{Results}
\label{sec:results}

\subsection{Cosmological constraints}
\label{sec:cosmo-par-constr}

In this section, we present the results from  constraining cosmological parameters of our Bianchi-I model for all four choices of anisotropic matter listed in Table~{\ref{tab:am-eos}} at the current epoch. We do four separate MCMC likelihood analysis for the four anisotropic sources considering one at a time along with the usual (isotropic) sources of the standard cosmological model.
Therefore, the set of cosmological parameters to be constrained are ${\boldsymbol\Theta}=\{{\boldsymbol\theta},M_0\}
=\{h_0, e^2_0, \sigma_0, \Omega_{AM}, \Omega_{IM},(l_a,b_a),M_0\}$.
We remind that $\Omega_\Lambda$ is fixed by the constraint equation following
Eq.~(\ref{eq:b1-constraint-eqn}).
Due to improved SNIa calibration techniques (quality) and a much larger data sample (quantity)
compared to previous releases~\cite{scolnic2018,betoule2014,scpsuzuki2012,perlmutter1999}, we get good constraints on all the parameters. The priors used for various parameters of the
models studied in the present work are listed in Table~\ref{tab:priors}.

\begin{table}
\centering
\begin{tabular}[t]{|| c | c ||}
\hline
\multicolumn{2}{||c||}{Bianchi-I} \\
\hline
$h_0$ & [0.5,1] \\
$e^2_0$ & [0,0.5] \\
$\sigma_0$ & [-0.2,0.2] \\
$\Omega_{AM}$ & [0,0.4] \\
$\Omega_{IM}$ & [0,0.5] \\
$l_a$ & $[0^\circ,360^\circ]$ \\
$b_a$ & $[-90^\circ,90^\circ]$ \\
$M_0$ & [-21,-18]\\
\hline
\end{tabular}
\quad\quad
\begin{tabular}[t]{ || c | c ||}
\hline
\multicolumn{2}{||c||}{$\Lambda$CDM} \\
\hline
$h_0$ & [0.5,1] \\
$\Omega_{M}$ & [0,0.5] \\
$M_0$ & [-21,-18]\\
\hline
\end{tabular}
\caption{Priors on various parameters of the model studied in the present work
         viz., anisotropic Bianchi-I model with one anisotropic matter source and the
         standard flat $\Lambda$CDM model for reference. The parameter `$\Omega_\Lambda$'
         is taken to be the dependant variable, automatically fixed following the constraint
         Eq.~(\ref{eq:b1-constraint-eqn}) for Bianchi-I model and via
         $\Omega_\Lambda=1-\Omega_{M}$ for the standard $\Lambda$CDM model.}
\label{tab:priors}
\end{table}

\begin{figure}
\centering
\includegraphics[width=0.49\textwidth]{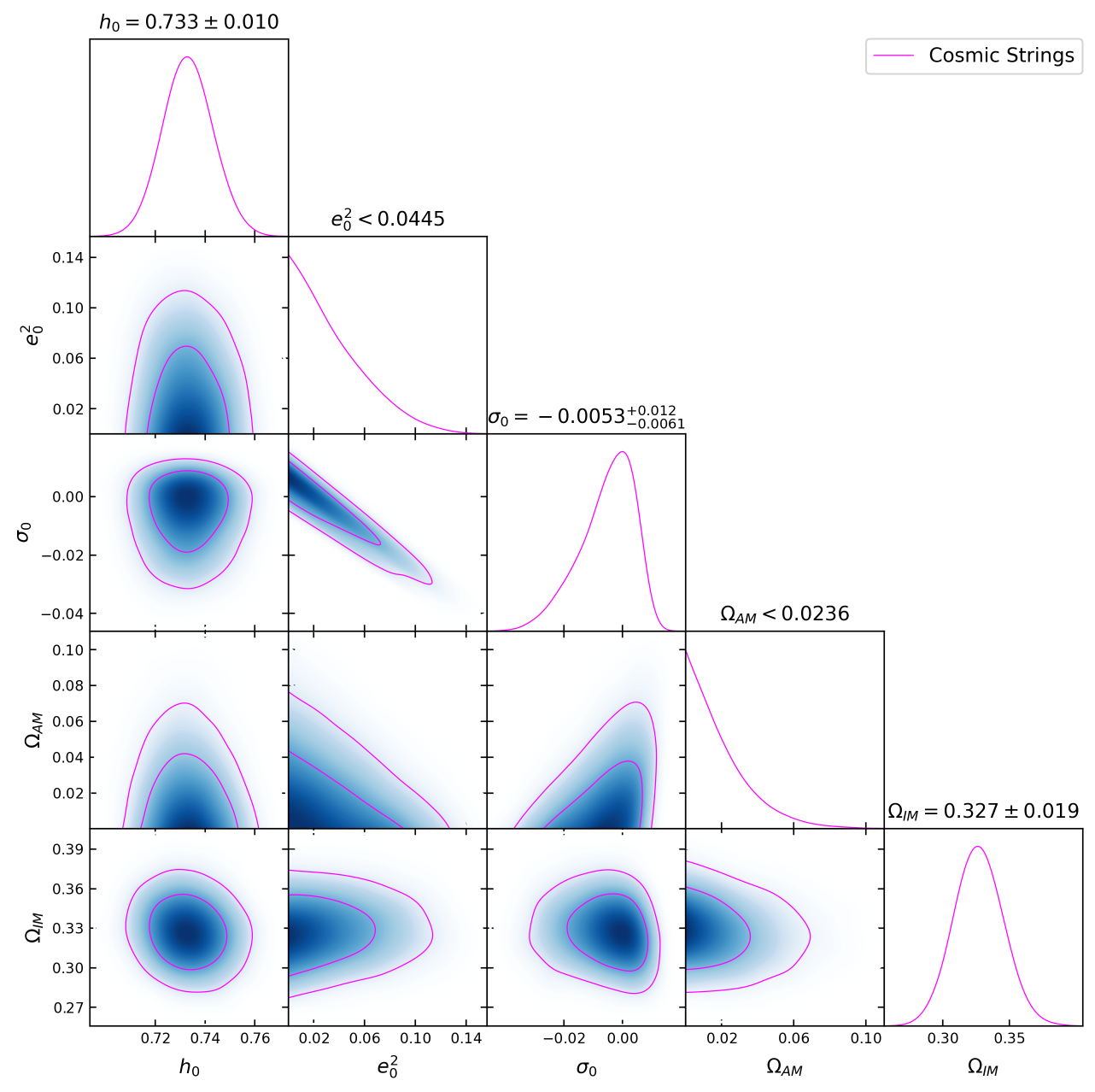}
~
\includegraphics[width=0.49\textwidth]{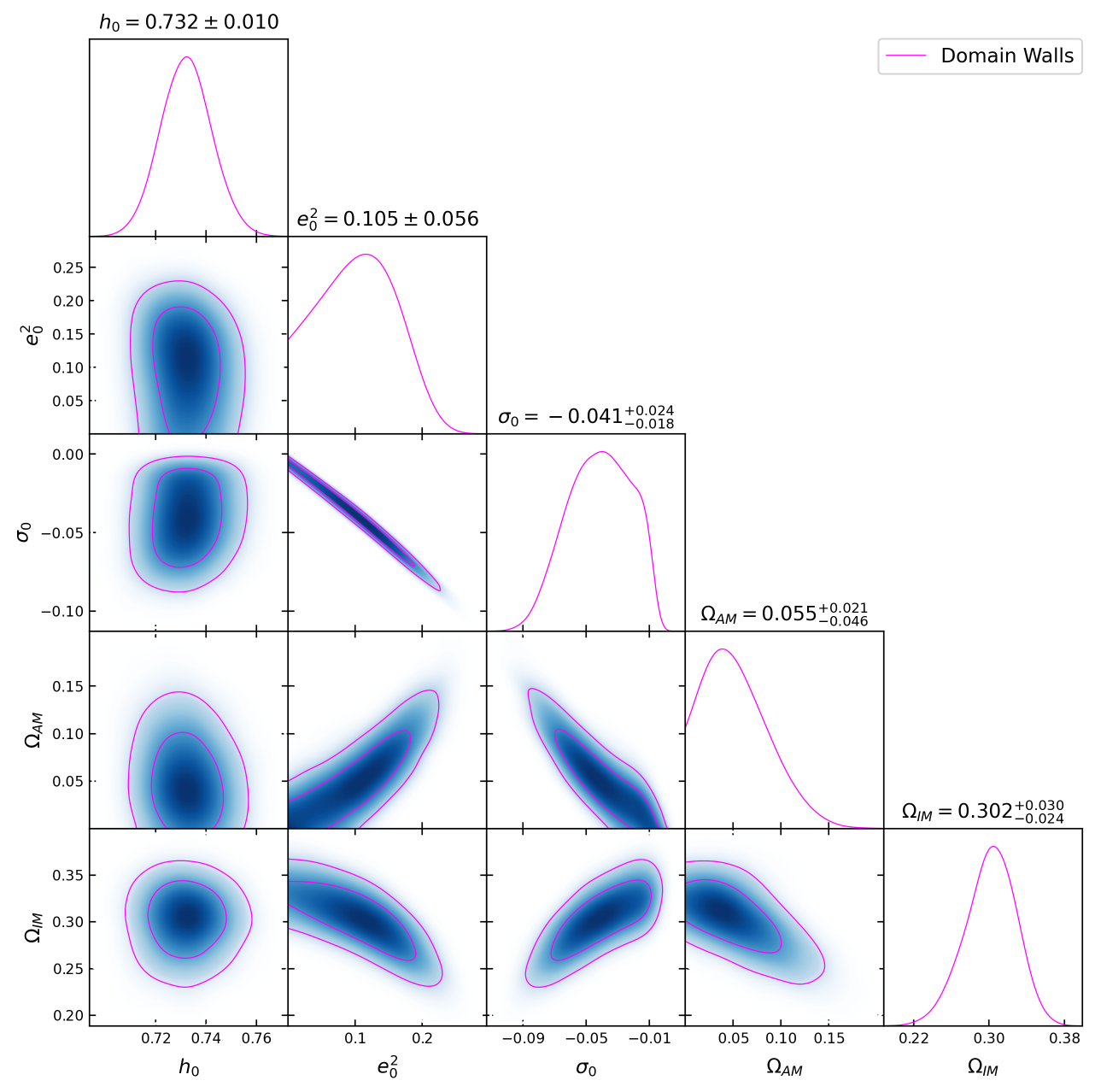}

\includegraphics[width=0.49\textwidth]{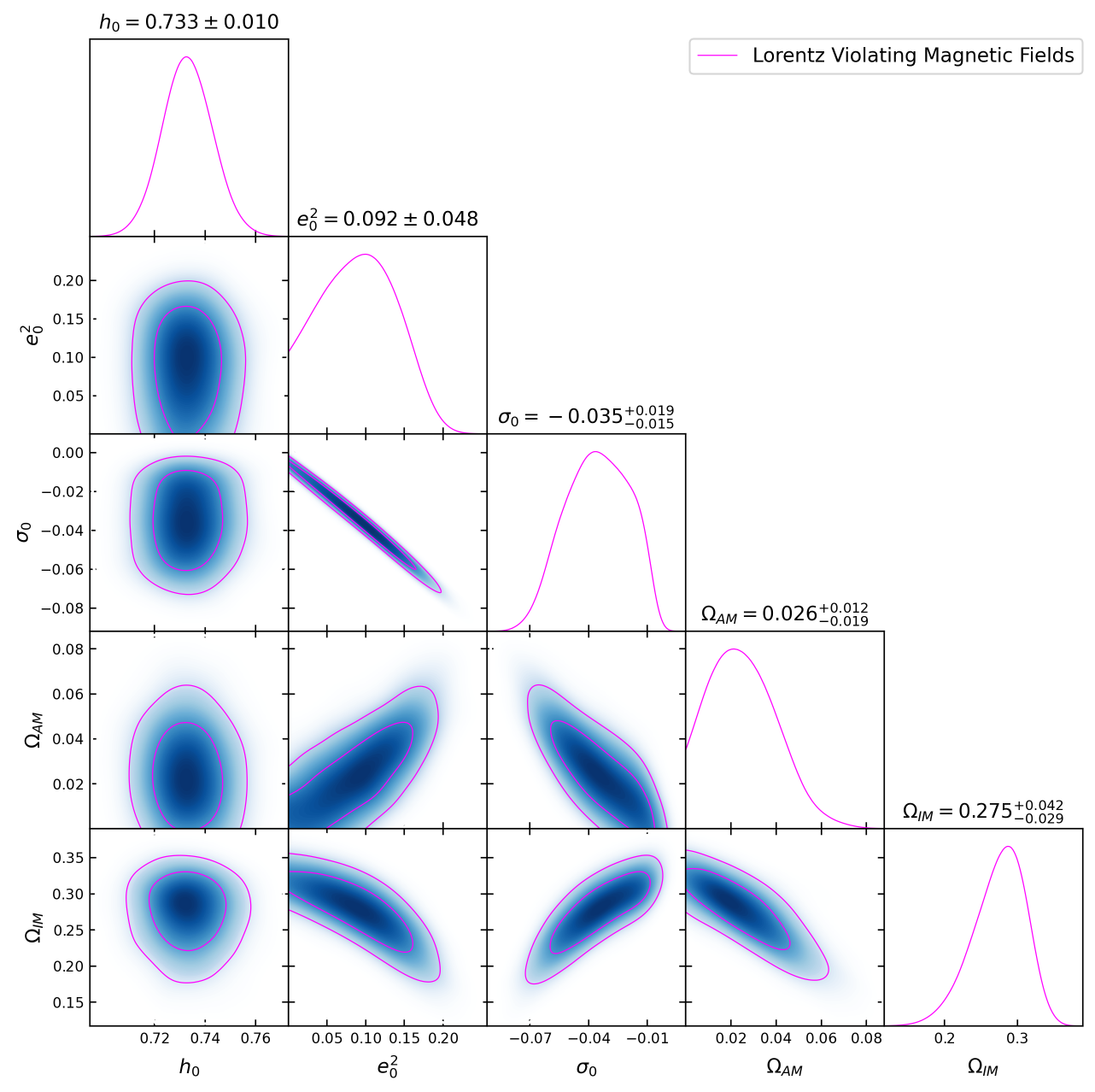}
~
\includegraphics[width=0.49\textwidth]{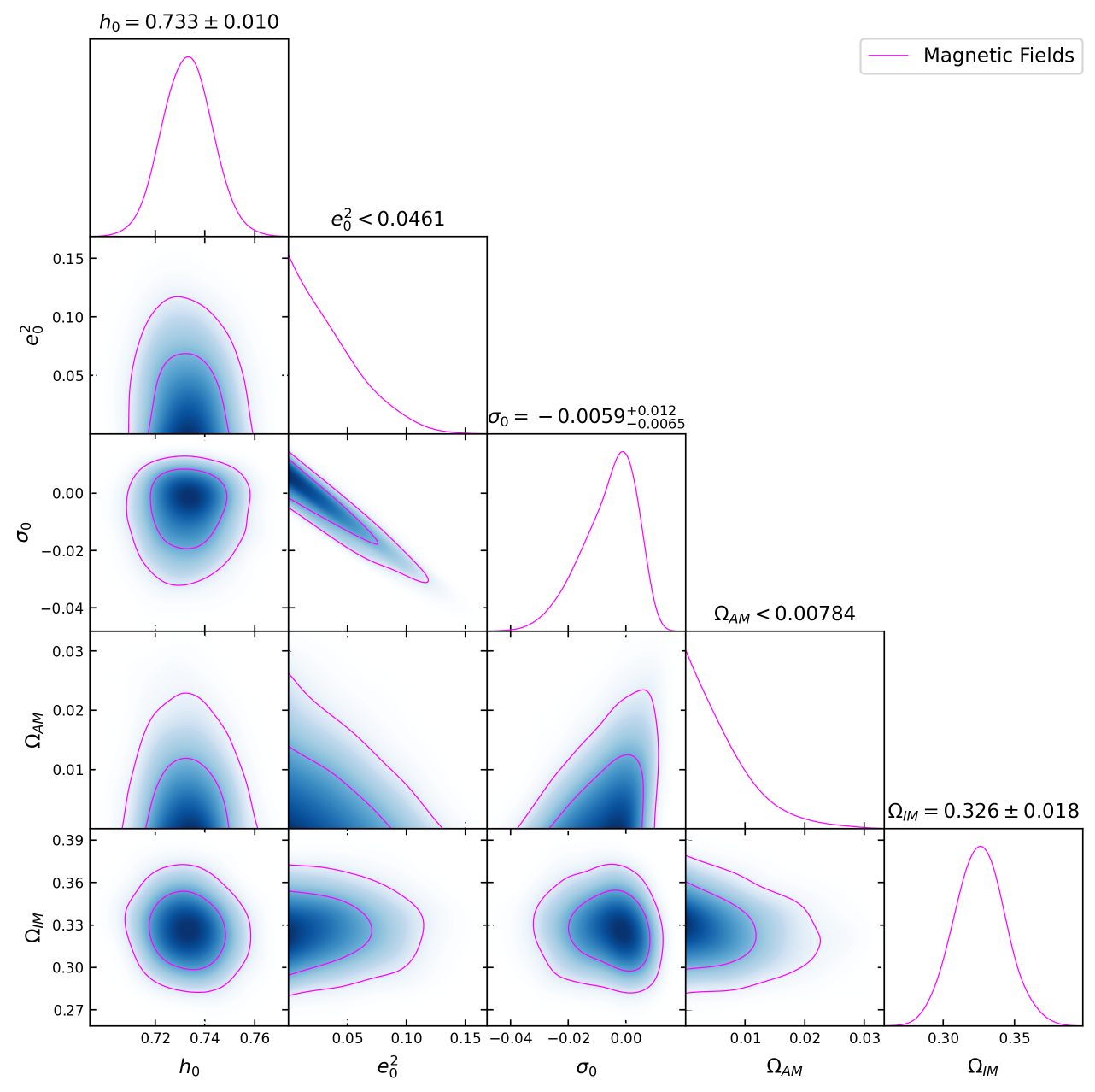}
\caption{2D contour plots of all Cosmological parameters in our Bianchi-I
		model assuming Cosmic Strings (CS), Domain Walls (DW), Lorentz
		Violation generated Magnetic Fields (LVMF), and Magnetic Fields (MF)
		as sources of anisotropic matter along with isotropic standard model
		sources.}
\label{fig:model-constr-mcmc}
\end{figure}

\begin{figure}
\centering
\includegraphics[width=0.45\textwidth]{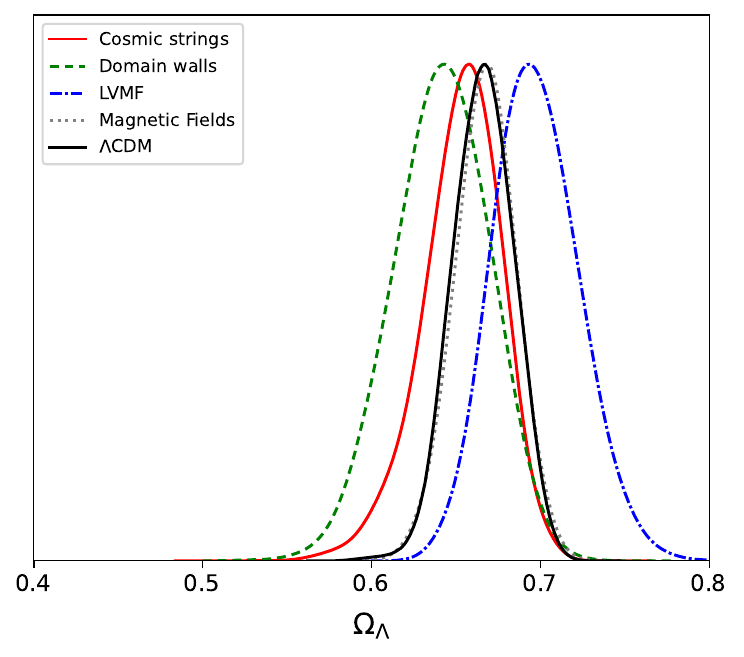}
~
\includegraphics[width=0.45\textwidth]{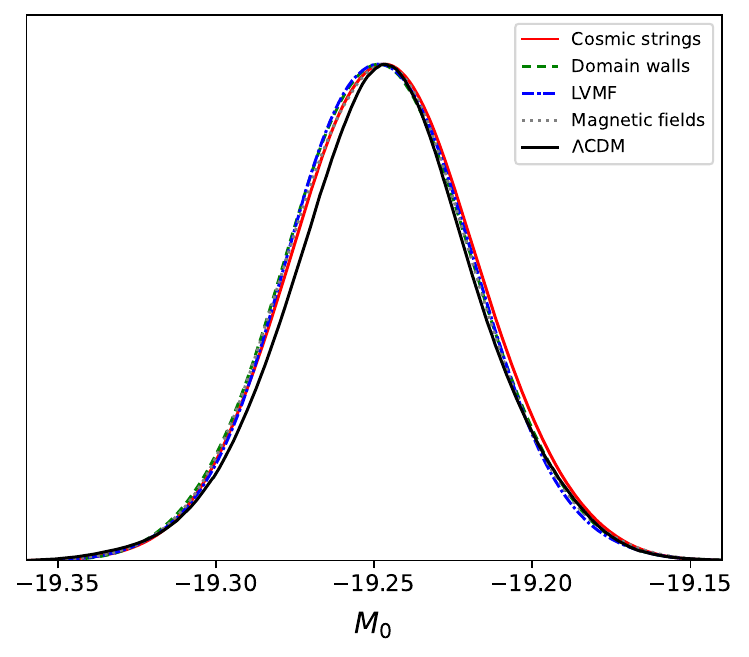}

\caption{\emph{Left} : Posterior distribution of dark energy density fraction
		($\Omega_\Lambda$) that is fixed by the constraint equation
		$\Omega_\Lambda = 1 - \Omega_{IM}-\Omega_{AM}-\sigma^2_0$ in case of
		anisotropic Bianchi-I model with one anisotropic matter type
		taken at a time, and using $\Omega_\Lambda=1-\Omega_{M}$ in case of
		flat $\Lambda$CDM model, where the others parameters are sampled as
		free parameters.
		\emph{Right} : Constraints obtained on the absolute B-band magnitude `$M_0$'
		of SNIa as obtained in both anisotropic Binahci-I model with different
		sources of anisotropy and the standard concordance model are shown
		here.}
\label{fig:ol_M0_all}
\end{figure}

Two dimensional contour plots with $1\sigma$ and $2\sigma$ confidence levels (CL) for all the
cosmological parameters with one anisotropic source included at a time viz., $\{h_0, e^2_0, \sigma_0, \Omega_{AM}, \Omega_{IM}\}$ are shown in Fig.~[\ref{fig:model-constr-mcmc}].
Following Eq.~(\ref{eq:b1-constraint-eqn}), we have chosen to sample all the parameters except for the cosmological constant term, $\Omega_{\Lambda}$.
The posterior distributions for `$\Omega_{\Lambda}$' from this constraint equation for all choices of anisotropic matter sources considered are shown in the \emph{left} panel of Fig.~[{\ref{fig:ol_M0_all}}].

\begin{figure}[!ht]
    \centering
    \includegraphics[width=0.54\textwidth]{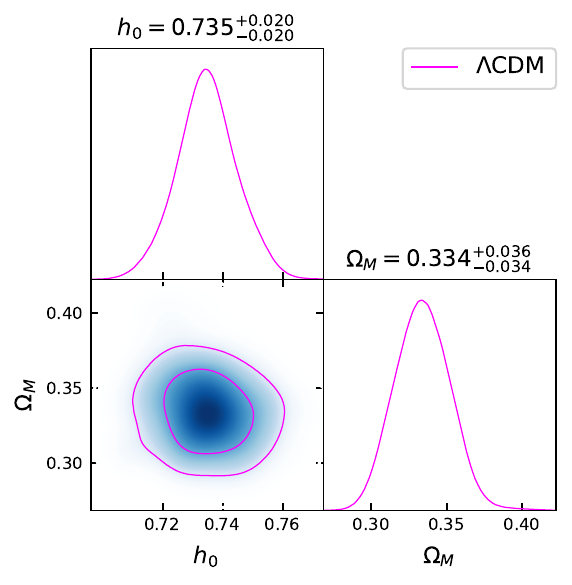}
    \caption{2D confidence contour plot for the flat $\Lambda$CDM model parameters
    $\{h_0,\Omega_{M}\}$ is shown here.}
    \label{fig:triplot_flrw}
\end{figure}

We also obtained constraints on the two parameter standard cosmological model i.e., the
$\Lambda$CDM model with a flat FLRW background, which is our base model to compare with.
Contour plots with $68\%$ and $95\%$ CL for the two parameter flat
$\Lambda$CDM model viz., \{$h_0,\Omega_{M}$\} are shown in Fig.~[\ref{fig:triplot_flrw}].
The posterior distribution of dark enegy denisty fraction `$\Omega_\Lambda$' in case of standard $\Lambda$CDM model following the constraint equation `$\Omega_\Lambda=1-\Omega_{M}$' is also shown in the \emph{left} panel of Fig.~[\ref{fig:ol_M0_all}]. We used very relaxed (uniform) priors on various parameters of the models studied here.

Constraints on the absolute B-band magnitude `$M_0$' were also obtained in both the case i.e., the anisotropic Bianchi-I model with different anisotropic matter sources and the standard flat $\Lambda$CDM model. As mentioned earlier, since we are using SH0ES calibrators (Cepheids), we were able to obtain simultaneous constraints on `$M_0$' and `$h_0$'. The constraints thus obtianed on `$M_0$' are shown in the \emph{right} panel of Fig.~[\ref{fig:ol_M0_all}]. As is obvious from the plot, we essentially get the same value for `$M_0$' in both models.

All the constraints obtained on various model parameters of anisotropic
Bianchi-I universe as well as the standard flat $\Lambda$CDM model are listed in
Table~\ref{tab:par-val-mcmc}.

\begin{sidewaystable}
 \centering
    \begin{tabular}{| c || c | c | c | c | c |}
    \hline
    Parameter & CS & DW & LVMF & MF & $\Lambda$CDM\\
    \hline
    $h_0$  & $0.733\pm0.010$ & $0.732\pm0.010$ & $0.733\pm0.010$ & $0.733\pm0.010$ & $0.735\pm{0.020}$\\[5pt]
    
    $\sigma_0$ & $-0.0053^{+0.012}_{-0.0061}$ & $-0.041^{+0.024}_{-0.018}$ & $-0.035^{+0.019}_{-0.015}$ & $-0.0059^{+0.012}_{-0.0065}$ & -\\[5pt]
    
    $e^2_0$ & $< 0.0445$ & $  0.105\pm{0.056}$ & $0.092\pm{0.048}$ & $<0.0461$ & -\\[5pt]
    
    $\Omega_{AM}$ & $< 0.0236$ & $0.055^{+0.021}_{-0.046}$ & $0.026^{+0.012}_{-0.019}$ & $< 0.00784$ & - \\[5pt]
    
    $\Omega_{IM}$ & $0.327\pm 0.019$ & $0.302^{+0.030}_{-0.024}$ & $0.275^{+0.042}_{-0.029}$ & $0.326\pm0.018$ & $0.334^{+0.036}_{-0.034}$\\[5pt]
    
    $\Omega_\Lambda$ & $0.653^{+0.026}_{-0.020}$ & $0.641^{+0.030}_{-0.027}$ & $0.697^{+0.023}_{-0.028}$ & $0.667\pm 0.018$ & $0.666^{+0.034}_{-0.036}$\\[5pt]
    
    $\hat{\lambda}=(l_a,b_a)$ & $(290.3^{+7.7}_{-36},9\pm17)$ & $(214^{+13}_{-25},-41.6^{+21}_{-6.7})$ & $(222^{+18}_{-28},-46^{+23}_{-9.8})$ & $(289.9^{+8.2}_{-34},10^{+13}_{-15}$ & -\\[5pt]
    
       $M_0$  & $-19.247\pm0.030$ & $-19.248\pm0.029$ & $-19.248\pm0.029$ & $-19.248\pm0.030$& $-19.247\pm{0.058}$\\[5pt]
    
    $\chi^2$ & 1523.7 & 1519.8 & 1520.1 & 1523.4 & 1525.3\\[5pt]
    
    $\Delta\chi^2$ & 1.6& 5.5& 5.2& 1.9& -\\[5pt]
    
    ${\rm log_{10}K}$ & -3.19 & -2.10 & -1.99 & -4.52  & 1\\
    \hline
    \end{tabular}
    \caption{Best fit values of all the cosmological parameters of anisotropic Bianchi-I
             model with cosmic strings (CS), domain walls (DW), Lorentz violation generated
             magnetic fields (LVMF) and magnetic fields (MF) considered as sources of
             anisotropy, including the axes of anisotropy in galactic coordinates, and the
             goodness-of-parameters are tabulated here.}
\label{tab:par-val-mcmc}
\end{sidewaystable}

From the likelihood density contour plots, one can see that various anisotropic sources
studied here led to two different patterns of constraints.
One resulted in non-zero anisotopic matter denisty ($\Omega_{AM}$) fraction
at present in case of Domain walls and LVMF.
On the other hand, constraints on anisotropic matter density fractions due to Cosmic
strings and Magnetic fields are consistent with zero.
Likewise, the constraints on cosmic shear ($\sigma_0$) at present times is also found
to be maximum, but negative and non-zero, in case of Domain walls and LVMF at a
$2\sigma$ confidence level.
However, in case of Cosmic strings and Magnetic fields as sources of anisotropy, the
cosmic shear is consistent with zero.
Note that the estimated cosmic shear parameter, $\sigma_0$, whether consistent with
zero or having a non-zero value, has a preference for a negative value for all choices of
anisotropic matter sources.
Given the fit for non-zero cosmic shear from SNIa data in case of Domain walls and
LVMF, we may infer that there may have existed a small fraction of energy density from
these sources in early universe which still remain inducing mild anisotropy at
current times.
We expect, then, that there will be similar pattern of constraints on the eccentricity paramter ($e_0^2$), i.e., a zero or non-zero constraint consistent with the corresponding anisotropy parameters $\sigma_0$ and $\Omega_{AM}$ for a particular source.
Indeed we find such a relation between them.
So there is an obvious correlation, expectedly, between the parameters
$\{e^2_0, \sigma_0, \Omega_{AM}\}$ that characterize the anisotropic nature of
a Bianchi-I space-time sourced by an anisotropic matter type.
We would like to note that the constratints on $e^2_0$ doesn't rule out `0' at
$2\sigma$ CL for Domain walls and LVMF also. This implies that $a_0=b_0=1$ i.e., the
scale factors are varying not very differently from each other in different directions
(within the errorbars) at present with the considered anisotropic matter sources. We
remind that the scale factors $a(t)$ and $b(t)$ can scale independently with the ratio
$a_0/b_0$ or vice versa not being equal to `1'.

\begin{figure}
    \centering
    \includegraphics[scale=0.72]{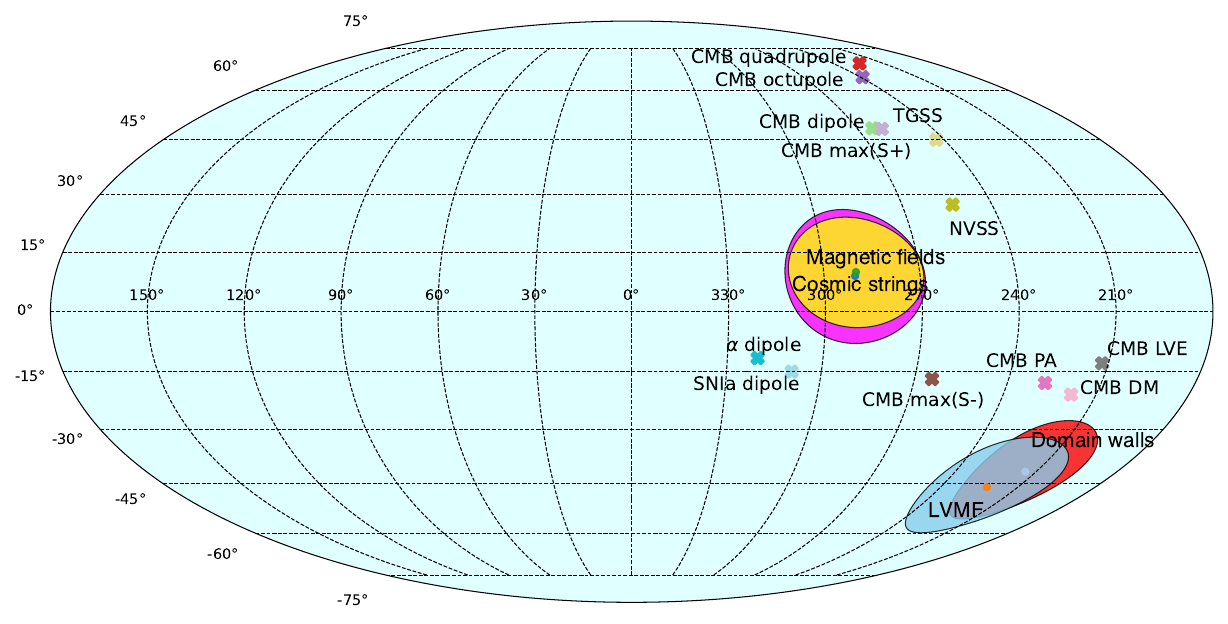}
    \caption{Mollweide view projection of the cosmic preferred axis for all four
             types of anisotropic matter considered in our work are shown with $1\sigma$
             confidence contour. Also shown are similar axis of anisotropy found
             in various astronomical/cosmological data listed in Table~\ref{tab:pref-axis}.}
    \label{fig:cosmic_axes}
\end{figure}

Another interesting observation is that the constraints on preferred axis for these
sources also follow a similar pattern. Similar anisotropy axes are inferred from Cosmic
strings and Magnetic fields, while Domain walls and LVMF lead to a different but
almost similar (overlapping) preferred axes.
All the four anisotropy axes found with each of the anisotropic matter source studied in the present work are shown in Fig.~[\ref{fig:cosmic_axes}] with $1\sigma$ CL contour.

The axis of anisotropy turns out to be $(l_a, b_a) \approx (290^\circ,9^\circ)$ and
$(290^\circ,10^\circ)$ for Cosmic strings and Magnetic fields, respectively, in galactic coordinates.
Also plotted along with these axes are some prominent anisotropy axes found in various data
sets in the upper right quadrant of the sky in galactic coordinates such as the CMB dipole ($\ell=1$) direction with which CMB quadrupole ($\ell=2$) and octupole ($\ell=3$) axes are also aligned~\cite{plk2013isostat}, the
radio dipole directions as found in NRAO VLA Sky Survey (NVSS) and TIFR GMRT Sky Survey (TGSS)~\cite{bengaly2018}. Recall that the bounds on current energy density fraction due to these two sources viz., Cosmic strings and Magnetic fields are consistent with zero.

With Domain walls and LVMF as sources of anisotropy, the axes of anisotropy were found to
be $(l_a,b_a) \approx (214^\circ,-42^\circ)$ and $(222^\circ,-46^\circ)$ respectively.
An agreement (albeit less precise) at least in the same quadrant can be seen for Domain walls and LVMF with various anisotropy axes seen in other data sets.
Specifically, the preferred axes found with Domain walls and LVMF as anisotropic matter sources broadly lie in the same direction as CMB hemispherical power asymmetry,
CMB cold spot, CMB mirror parity asymmetry axis, etc., which are all some of the significant
anomalous anisotropy axes or features seen in CMB data~\cite{plk2015isostat}. Here we remind ourselves that the constraints on energy density fractions on Domain walls and LVMF prefer a non-zero value at $2\sigma$ CL.

Note that in a previous analysis~\cite{aluri2013sn1a} using Union 2 data~\cite{amanullah2010}, the axis of anisotropy for Cosmic strings, Magnetic fields,  Domain walls and LVMF were all found to be broadly well aligned (in the lower right quadrant of the celestial sphere in galactic coordinates).

So, we can see that there are two sets of cosmic axes broadly spanning the upper right and lower right quadrants in Fig.~\ref{fig:cosmic_axes}.
Those that are pointing in the CMB dipole direction (or broadly in its vicinity), may possibly be arising due to an unknown systematic or data processing artifact, if not a cosmic preferred axis, as is the case with anisotropy axes found for Cosmic strings and Magnetic fields. It would be reasonable to say so given that we were only able to obtain upper bounds on the energy density fractions in case of Cosmic strings and Magnetic fields as anisotropic matter sources. However the anisotropy axes seen when assessing Domain walls and LVMF as sources of anisotropy could be an indication of a real cosmic preferred axis for our Universe, in whose case we found a non-zero energy density fraction at 95\% CL. This reasoning is further reinforced by the fact that other anisotropy axes in their proximity are all
significant large scale CMB anomalies among others.

\begin{table}[t]
\begin{center}
    \begin{tabular}{|c | c | c |}
    \hline
    Axis & $(l_a,b_a)$ & Ref.\\
    \hline
    CMB dipole & (263.99,48.26) & \cite{plk2018maps} \\
    CMB quadrupole & (224.2,69.2) & \cite{plk2013isostat}\\
    CMB octupole & (239,64.3) & \cite{plk2013isostat}\\
    CMB Even mirror parity axis & (260.4,48.1) & \cite{plk2015isostat}\\
    CMB Odd mirror parity axis & (264.4,-17) & \cite{plk2015isostat}\\
    CMB Hemispherical Power Asymmetry axis & (218,-21) & \cite{plk2015isostat}\\
    CMB Dipole Modulation axis& (228,-18) & \cite{plk2015isostat}\\
    CMB Local Variance estimator& (212,-13) & \cite{akrami2014}\\
    Radio dipole NVSS& (253.12,27.28) & \cite{bengaly2018}\\
    Radio dipole TGSS& (243,45) & \cite{bengaly2018}\\
    Fine structure constant ($\alpha$) dipole & (320.5,-11.7) & \cite{alpha_dipole}\\
    SNIa dipole& (316,-5) & \cite{SNIa_dipole}\\
    \hline
    \end{tabular}
    \caption{Other cosmic preferred axes (in galactic coordinates), depicted in      
             Fig.~[\ref{fig:cosmic_axes}], as found in various astronomical/cosmological
             data are listed here.}
\label{tab:pref-axis}
\end{center}
\end{table}

\subsection{Goodness of fit}
\label{sec:bayes-factor-comparison}
While performing any kind of `goodness of fit' test, two different models used to fit the same data having different number of fit parameters may seem difficult to compare, both qualitatively and quantitatively. One approach would be to simply check the difference in the minima of the $\chi^2$ for the two models wherein one of the models is taken as base model, and also calculating the $\chi^2$ per degree of freedom. For the various (alternate) models considered in the present work, the values of $\Delta\chi^2=\chi^2_{\rm ref} - \chi^2_{\rm alt}$ are listed in the second row from the bottom of Table~\ref{tab:par-val-mcmc}, with respect to the (reference) flat $\Lambda$CDM model.

Introducing more parameters to describe observed data almost always leads to a better fit i.e., it returns a lower $\chi^2$ minimum value. So a superior procedure in identifying the right model that describes the observed data should factor in the additional parameters introduced into an \emph{alternate} model with respect to a \emph{reference} model and
also the parameter space (prior volume).

One such model comparison method is Bayes approach. Central to Bayesian inference schemes
is the Bayes theorem which states that, given a data vector `${\bf d}$' that is being
described by a model $M$ with parameters ${\boldsymbol \theta}$, then,
\begin{equation}
P({\boldsymbol \theta}|{\bf d}) =
\frac{P({\bf d}|{\boldsymbol \theta})P({\boldsymbol \theta})}{P({\bf d})}\,,
\end{equation}
where $P({\boldsymbol \theta}|{\bf d})$ is the posterior distribution of the model
parameters given the data, $P({\bf d}|{\boldsymbol \theta})=\mathcal{L}({\boldsymbol \theta})$ is the likelihood function of the data given the parameters ${\boldsymbol \theta}$ (i.e., under the assumption of model $M$), $P({\boldsymbol \theta})=\pi({\boldsymbol \theta})$ is our prior knowledge (range) of the model parameters, and finally $P({\bf d})=Z$ is called the Bayesian evidence.
The evidence `$Z$' is essentially the marginalized likelihood over the parameters prior volume, that is also the normalization factor, which serves as a measure of how well a model explains the observed data while penalizing for model complexity. The evidence encapsulates the plausibility of the assumed model in adequately describing the data at hand, integrated over the entire parameter space, providing a principled basis for model selection.
As such, it is given by,
\begin{equation}
Z = \int \mathcal{L}({\boldsymbol \theta}) \pi({\boldsymbol \theta}) d{\boldsymbol \theta}\,.
\end{equation}

In this study, we employ the Bayes Factor `$K$' as a comparative metric to compare an alternative model with a reference model. It is defined as the ratio of the evidences of the two models $Z_{\rm alt} / Z_{\rm ref}$. Evidence is calculated using (dynamic) Nested sampling approach~\cite{skilling2004,skilling2006,higson2019}. Numerically, one obtains the logarithm of the marginal likelihoods i.e., logarithm of evidence, $\ln(Z)$. Thus, the Bayes factor is defined as,
\begin{equation}
    K = e^{\{\ln (Z_{\rm alt}) - \ln (Z_{\rm ref})\}}\,.
\end{equation}
To facilitate interpretation, we express the Bayes factor `$K$' in logarithmic form,
$\log_{10}K$, enabling straightforward inference based on Jeffreys' scale which is an established convention for guaging evidence strength~\cite{jeffreys1998book,trotta2008}.
A summary of evidence strength depending on the Bayes factor is given in
Table~\ref{tab:bayes-evidence}.

\begin{table}[h]
\centering
\begin{tabular}{|c|c|c|c|}
\hline
\text{$\log_{10}K$} & \text{Odds} & \text{Probability} & \text{Strength of evidence} \\
\hline
$< 1.0$ & $\lesssim 3:1$ & $< 0.750$ & Inconclusive \\
1.0 & $\sim3:1$ & 0.750 & Weak evidence \\
2.5 & $\sim12:1$ & 0.923 & Moderate evidence \\
5.0 & $\sim150:1$ & 0.993 & Strong evidence \\
\hline
\end{tabular}
\caption{Strength of Evidence based on $\log_{10} K$}
\label{tab:bayes-evidence}
\end{table}

The Bayes factor i.e., $\log_{10} K$, for various models studied in the present work are listed in the last row of Table~\ref{tab:par-val-mcmc}.
As we can see from that table, the evidence to consider various anisotropic
sources is \emph{inconclusive} over the base flat $\Lambda{CDM}$ model.

However, given the non-zero constraints obtained on anisotropic energy density fractions in case of Domain walls and LVMF at $2\sigma$ CL, perhaps they cannot be entirely negated. Future
data sets with much larger sample sizes may allow us to obtain definitive constraints on these
sources.

\section{Conclusions}
\label{sec:concl}
Our goal here, assuming a Bianchi-I universe, was to estimate the level of anisotropy in the `observed' universe by estimating the cosmic shear and eccentricity of the universe, constrain the density fractions of (an)isotropic matter sources and dark energy modeled as cosmological constant ($\Lambda$) at current times, and identify any cosmic preferred axis for our universe. We considered various anisotropic matter sources, specifically, Cosmic strings, Magnetic fields, Domain walls and Lorentz violation generated magnetic fields (LVMF) in the source term of Einstein's equation, taken one at a time, along with the standard model (isotropic) sources viz., dust-like (dark+visible) matter and dark energy ($\Lambda$). Bianchi-I model that we studied here is an anisotropic cosmological model with a planar symmetry in the spatial part of it's metric (as defined in Eq.~\ref{eq:bianchi1plnr}). The evolution equations corresponding to various cosmological parameters of this model in the form of coupled differential equations (Eq.~\ref{eq:b1evoleq}) were rederived. We then performed an MCMC likelihood analysis of the Bianchi-I model solving these evolution equations in conjugation with the corresponding luminosity distance vs redshift relation (Eq.~\ref{eq:b1lumdist}) by defining a $\chi^2$ function (Eq.~\ref{eq:chi2-cov}). We made use of the latest compilation of Type Ia Supernova (SNIa) data from the \emph{Pantheon}+ collaboration~\cite{scolnic2022}.

The current work is an extension of the analysis presented in Ref.~\cite{aluri2013sn1a} where constraints on a Bianchi-I universe with the same anisotropic matter sources were obtained using Union2 SNIa data~\cite{amanullah2010}, to probe for any signatures of isotropy violation in the universe. With nearly thrice the data size in this Pantheon+ compilation compared to Union2 sample, we get better constraints on the parameters of the anisotropic Bianchi-I model.

With Pantheon+ data set, we found that the four anisotropic sources considered here broadly lead to two patterns of similar parameter constraints. Cosmic strings and Magnetic fields resulted in nearly same parameter values, and Domain walls and LVMF were found to have different but similar parameter values.
We obtained non-zero constraints for the cosmological parameters describing anisotropic universe when considering Domain walls and LVMF as sources of anisotropy.
There is a preference for a negative cosmic shear,
non-zero eccentricity parameter, and non-zero anisotropic matter denisty fraction at
95\% CL at present.
In case of Cosmic strings and Magnetic fields, constraints on cosmic shear, eccentricity parameter and anisotropic matter density fraction are consistent with zero.

Also the cosmic anisotropy axes found here fall in two distinct regions.
In case of Cosmic strings and Magnetic fields, whose anisotropic cosmological parameters viz., $(e^2_0, \sigma_0, \Omega_{AM})$ were found to be consistent with zero, their axes of anisotropy broadly lies in the upper right quadrant as the CMB dipole ($\ell=1$). We suspect that this may be due to some systematic in the data, which is inducing to pick up the direction found here. In case of Domain walls and LVMF, their axes of anisotropy reported are broadly aligned with other prominent axes of isotropy violation such as seen in CMB data in the lower right quadrant of the sky in galactic coordinates.

The (mean) Hubble parameter value ($h_0$) we obtained in this Bianchi-I model analysis with various anisotropic sources doesn't elleviate the `Hubble tension'~\cite{valentino2021,efstathiou2020,freedman2021}.
Thus, we may say that the `Hubble tension' still remains unresolved with this model.

Finally, when we assess the evidence for a particular model/anisotropic source to likely describe the data well, we find that a flat $\Lambda$CDM model is still preferable.
However, a Bianchi-I model with Domain walls and LVMF as anisotropic matter sources may still be considered further as their anisotropic matter density constraints are non-zero at $2\sigma$ CL and the corresponding anisotropy axes are broadly aligned with some of the well known isotropy violating axes found in CMB sky. We therefore conclude that more data may be needed to arrive at stringent constraints to decide on these alternate anisotropic models.

\section*{Acknowledgements}

AV acknowledges the financial support received through research fellowship awarded by Council of Scientific \& Industrial Research (CSIR), India during this project. We acknowledge National Supercomputing Mission (NSM) for providing computing resources of ‘PARAM
Shivay’ at Indian Institute of Technology (BHU), Varanasi, which is implemented by C-DAC and supported by the Ministry of Electronics and Information Technology (MeitY) and Department of Science and Technology (DST), Government of India. DFM thanks the Research Council of Norway for their support and the resources provided by UNINETT Sigma2 - the National Infrastructure for High Performance Computing and Data Storage in Norway.
The authors would like to thank David Jones and Tamara Davis for many helpful communications regarding Pantheon+ data.
We acknowledge using \texttt{Cobaya}\footnote{\url{https://cobaya.readthedocs.io/en/latest/}}  \cite{cobaya2021} - a code for Bayesian analysis in Cosmology, \texttt{dynesty}\footnote{\url{https://dynesty.readthedocs.io/en/stable/}} - a Dynamic Nested Sampling package for estimating Bayesian posteriors and evidences~\cite{dynesty,dynest-zenodo}, and \texttt{GetDist}\footnote{\url{https://getdist.readthedocs.io/en/latest/}} \cite{getdist2019} - a user friendly GUI interface for the analysis and plotting of MCMC samples. We also acknowledge the use of \texttt{SciPy}\footnote{\url{https://scipy.org/}}~\cite{scipy2020}, \texttt{NumPy}\footnote{\url{https://numpy.org/}}~\cite{numpy2020}, \texttt{Astropy}\footnote{\url{http://www.astropy.org}}~\cite{astropy2013,astropy2018,astropy2022}, and \texttt{Matplotlib}\footnote{\url{https://matplotlib.org/}}~\cite{hunter2007}. The authors thank the anonymous referee for
constructive comments and insightful suggestions.

\bibliographystyle{unsrt}
\bibliography{ref_av1}

\end{document}